%% file: main.tex
\documentclass[%
 reprint,
 amsmath,amssymb,
 aps,
prb,
]{revtex4-2}

\usepackage{lipsum}
\usepackage{graphicx}
\usepackage{dcolumn}
\usepackage[caption=false]{subfig}
\usepackage{bm}
\usepackage{hyperref}
\hypersetup{
    colorlinks=true,
    linkcolor=blue,
    filecolor=magenta,      
    urlcolor=cyan,
    citecolor = blue,
    }
\usepackage{physics}
\usepackage{tikz}
\usetikzlibrary{arrows.meta}
\usepackage{amsmath}

\usepackage{comment}
\usepackage{enumerate}
\usepackage{float}
\usepackage{multirow}
\usepackage{orcidlink}

\begin{document}

\preprint{APS/123-QED}

\title{Langmuir kinetics and totally asymmetric volume exclusion in linear chains \\via Fock space formalism}

\author{Luis D. S. Silva
\orcidlink{0009-0008-5099-4767}}
\email{ldavifisico@gmail.com}
\affiliation{%
 Departamento de Física, Universidade Federal de Sergipe, 49107-230 São Cristóvão, Sergipe, Brazil.}%

\author{Arthur M. S. Carvalho \orcidlink{0009-0009-5279-1290}}
\email{arthurcarvalho.ufs@gmail.com}
 \affiliation{%
 Departamento de Física, Universidade Federal de Sergipe, 49107-230 São Cristóvão, Sergipe, Brazil.}%
\author{Francisco A. G. Almeida \orcidlink{?}}%
\affiliation{%
 Departamento de Física, Universidade Federal de Sergipe, 49107-230 São Cristóvão, Sergipe, Brazil.}%
\author{Gerson C. Duarte-Filho \orcidlink{0000-0003-4975-4981}}%
\affiliation{%
 Departamento de Física, Universidade Federal de Sergipe, 49107-230 São Cristóvão, Sergipe, Brazil.}%


\date{\today}

\begin{abstract}

Driven-diffusive systems combining spatial exclusion and bulk exchange, such as the totally asymmetric simple exclusion process coupled with Langmuir kinetics (TASEP-LK), serve as foundational models for particle transport in vehicular traffic, conveyor networks, and mesoscopic flow channels. In this work, we employ an exact Doi-Peliti field-theoretic operator formulation in Fock space to evaluate steady-state observables, continuous phase landscapes, and spatial correlations in finite linear chains up to $k = 16$. By introducing a decomposition in terms of biorthonormal eigenvectors for asymmetric non-Hermitian matrices, the stationary state is calculated directly from the kernel of the quasi-Hamiltonian, completely bypassing the computationally prohibitive complexity of full matrix diagonalization and time-dependent matrix exponentiation. Across distinct rate regimes and global parametric scans, our exact discrete calculations demonstrate that finite-size stochastic fluctuations naturally smear out singular domain walls, producing smooth density profiles and continuous phase crossovers. Finite-size scaling illustrates how local profiles systematically steepen toward the sharp domain wall predicted by continuum field theory as $k \to \infty$. Furthermore, two-point correlation analysis reveals that in slow-kinetics regimes, spatial fluctuations generate strong inter-site cumulants, causing standard mean-field approximations to systematically underestimate joint occupancy probabilities by 50\%. These findings emphasize the necessity of analytical methods for modeling mesoscopic transport systems and short traffic corridors where local stochastic fluctuations and spatial correlations dominate the stationary dynamics.

\end{abstract}

\maketitle

\section{Introduction}

Stochastic processes underlie a wide range of physical phenomena, encompassing classical examples such as Brownian motion as well as more intricate systems, including epidemic propagation and intracellular transport \cite{kampen, redner}. The systematic investigation of these processes traces back to Robert Brown’s 1827 observations of the erratic motion of pollen grains suspended in water \cite{Brown}, while the rigorous theoretical framework was subsequently developed through the seminal contributions of Einstein and Smoluchowski \cite{Einstein, SMOL}. 

Among the wide range of stochastic phenomena, diffusion processes on one-dimensional lattices (linear chains) are of particular interest \cite{tasep,Voll}. The Totally Asymmetric Simple Exclusion Process (TASEP), introduced in 1968 \cite{tasepofc,Derrida,grune,Kuniba2022,Essler,Alonso2024robot}, describes unidirectional particle transport subject to hard-core exclusion \footnote{Hard-core exclusion refers to an interaction between particles modeled as impenetrable objects, such that a given site cannot be occupied by more than one particle simultaneously.}, where particles move forward provided the target site is vacant.
On the other hand, Langmuir kinetics (LK), formulated by Irving Langmuir in 1918 \cite{lk,islam,ergun}, provides a theoretical framework for adsorption-desorption dynamics at surfaces. Hybrid TASEP-LK models have been successfully employed to describe a variety of nonequilibrium transport processes \cite{Pal2026finite, Sato2016relaxation, Baiesi2026cycles, Gupta2025exclusion}, ranging from intracellular biological transport \cite{Bressloff, Panman} and protein synthesis along messenger RNA \cite{sharma} to macroscopic vehicular traffic flow \cite{CHOWDHURY2000199, Wassmer2026, Schadschneider_1997}.

To gain physical intuition into such driven many-body systems, the TASEP with LK can be naturally interpreted through the lens of vehicular traffic engineering. In this conceptual framework, the one-dimensional lattice acts as a single-lane highway where particles represent cars moving forward under a strict no-overtaking constraint, embodying the core exclusion principle of TASEP. The attachment and detachment of particles along the bulk correspond, respectively, to entry points (on-ramps) and exits to secondary roads (off-ramps), capturing the essential mechanisms of LK dynamics. Within this picture, a sudden spatial transition from a low-density (free-flow) regime to a high-density (jammed) regime represents a traffic shock wave, which mathematically manifests as a Domain Wall (DW).

Traditional analytical and numerical approaches to these non-equilibrium steady states typically rely on Mean-Field Approximations (MFA) or Monte Carlo simulations (MCS) \cite{parmegianni2003,parmegianni2004}. However, MFA systematically neglects spatial correlations, while MCS can suffer from statistical fluctuations and operational stiffness \cite{samant2005overcoming}, especially in mesoscopic systems. Consequently, there remains a clear need for non-perturbative exact frameworks capable of capturing finite-size fluctuations and spatial correlations in driven lattice systems without relying on mean-field assumptions.

In this work, we address these limitations by investigating non-equilibrium particle transport in a finite one-dimensional lattice of size $k$ governed by the TASEP with LK diffusion. By formulating the underlying Master Equation in Fock space \cite{Doi1,Doi2,Fock,Baez}, we deploy an operator methodology \cite{gerson,Santos,arthur_artigo,Knebel,Sato2016relaxation} 
based on a biorthonormal eigenvector decomposition of the non-Hermitian quasi-Hamiltonian to derive exact expressions for steady-state observables, including local density profiles and parametric mapping. We systematically explore six distinct physical transport regimes, explicitly demonstrating how mesoscopic finite-size corrections reshape phase boundaries and domain wall profiles. Our exact results provide a rigorous benchmark that exposes the breakdown of MFA in small- to intermediate-sized lattices, while avoiding the finite-time and finite-ensemble sampling artifacts inherent to Monte Carlo simulations.

Specifically, our exact mesoscopic calculations reveal that finite-size stochastic fluctuations naturally smear out sharp field-theoretic singularities into continuous density profiles. Furthermore, global parametric mapping demonstrates that incorporating Langmuir kinetics into the TASEP framework substantially broadens the intermediate region (\textbf{IR}) of moderate densities. Interpreted through traffic dynamics, this pronounced expansion provides a wide operational window where mesoscopic transit corridors maintain an optimal, stable throughput without suffering from starvation or severe congestion.

The structure of this paper is organized as follows. Sec.~\ref{sec:tasep+lk} examines the TASEP phenomena coupled with LK on one-dimensional linear chains, with additional emphasis on applications motivated by traffic engineering. Sec.~\ref{sec:second} presents a second-quantization formulation in terms of creation and annihilation ladder operators, whereby the master equation is mapped onto a Schrödinger-type evolution equation; this framework is then used to derive the explicit quasi-Hamiltonian operator governing the dynamics, along with the computation of steady-state observables via biorthonormal eigenvector decomposition. In Sec.~\ref{sec:results}, we report the main results for finite chains. We specifically highlight how finite-size fluctuations smooth out domain wall discontinuities, how Langmuir kinetics expands optimal traffic corridors, and how spatial correlations lead to the breakdown of mean-field predictions. Finally, concluding remarks and future perspectives are summarized in Sec.~\ref{sec:conclusion}.

\input{sec2}

\input{results}
\input{conclusion}

\begin{acknowledgments}
This work was supported by the Coordenação de aperfeiçoamento de Pessoal de Nível Superior (CAPES), Grant No. 88887.201072/2025-00. 
\end{acknowledgments}


\bibliography{bib}

\end{document}

%% file: sec2.tex
\section{TASEP and LK}
\label{sec:tasep+lk}

To model non-equilibrium transport in one-dimensional systems, we consider the TASEP coupled with LK on a discrete linear chain of $k$ sites. The system dynamics strictly enforce hard-core exclusion, meaning that each lattice site $j \in \{1, \dots, k\}$ can be occupied by at most one particle at any time, $n_j \in \{0,1\}$. 

\begin{figure}[htpb]
    \centering
    \resizebox{\columnwidth}{!}{
        
    \begin{tikzpicture}[
        >=Stealth,
        quadrado/.style={minimum size=0.75cm, inner sep=0pt},
        seta/.style={->, line width=0.8pt},
        rotulo/.style={font=\small}
    ]

    \foreach \x/\cor in {
        0/magenta!30,
        1.15/teal!25,
        2.30/teal!25,
        3.45/teal!25,
        4.60/teal!25,
        5.75/teal!25,
        6.90/teal!25,
        8.05/teal!25,
        9.20/magenta!30
    }{
        \fill[\cor] (\x,0) rectangle ++(0.75,0.75);
    }

    \fill[orange!80] (0.38,1.3) circle (0.19);

    \draw[seta] (0.38,1.05) -- (0.38,0.35);
    \node[rotulo] at (0.6,0.85) {$\alpha$};

    \fill[orange!80] (6.13,1.3) circle (0.19);

    \draw[seta] (6.12,1.05) -- (6.12,0.35);
    \node[rotulo] at (6.45,0.85) {$\omega_A$};

    \fill[red!65] (2.68,0.38) circle (0.20);
    \draw[white, line width=1pt] (2.68,0.38) circle (0.20);

    \draw[seta] (2.95,0.38) -- (3.95,0.38);
    \node[rotulo] at (3.28,0.78) {$D$};

    \draw[seta] (2.68,0.1) -- (2.68,-0.65);
    \node[rotulo] at (2.98,-0.15) {$\omega_D$};

    \fill[red!65] (9.58,0.38) circle (0.20);
    \draw[white, line width=1pt] (9.58,0.38) circle (0.20);

    \draw[seta] (9.6,0.1) -- (9.6,-0.65);
    \node[rotulo] at (9.8,-0.18) {$\beta$};

    \end{tikzpicture}
    }
    \caption{Schematic representation of the TASEP-LK dynamics on a linear lattice of $k$ sites, interpreted as a single-lane vehicular transport system. Magenta squares denote boundary sites, while teal squares represent the bulk. Particles (red circles on the lattice, orange circles in the reservoir) hop to the right at rate $D=1$ subject to hard-core exclusion (no overtaking). Boundary injection at rate $\alpha$ and ejection at rate $\beta$ represent the main highway entrance and exit, respectively. Bulk Langmuir kinetics accounts for particle attachment at rate $\omega_A$ (\emph{on-ramps}) and detachment at rate $\omega_D$ (\emph{off-ramps}).}
    \label{fig:tasep+lk}
\end{figure}
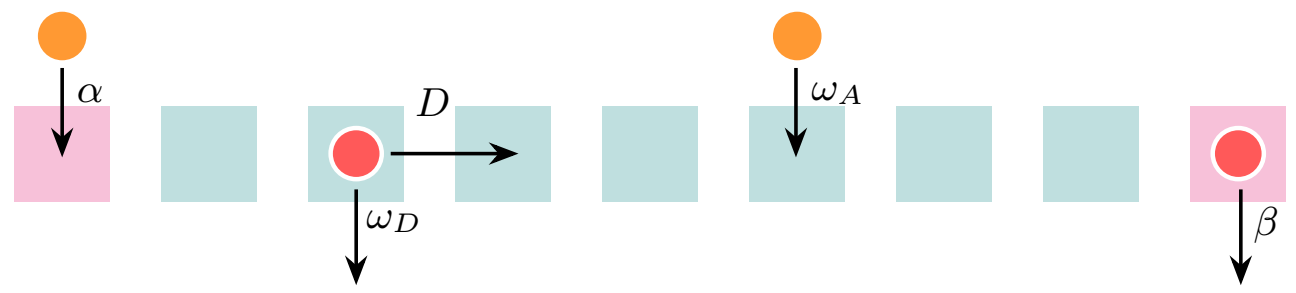

As illustrated schematically in Fig.~\ref{fig:tasep+lk}, to bridge statistical mechanics with transport engineering, this driven many-body system can be intuitively interpreted as a single-lane highway. Particles correspond to vehicles moving unidirectionally from left to right under a strict no-overtaking rule. Particles hop to an adjacent vacant site to the right at a normalized hopping rate $D = 1$, representing free-flow vehicular movement. The lattice is coupled at its boundaries to particle reservoirs: at the leftmost site ($j=1$), particles enter the chain at rate $\alpha$ (main highway entrance), provided the site is vacant; at the rightmost site ($j=k$), particles exit the system at rate $\beta$ (main highway exit). Langmuir Kinetics acts on all bulk sites ($j = 2, \dots, k-1$), modeling localized exchanges with external reservoirs. Vacant bulk sites undergo particle attachment (adsorption) at rate $\omega_A$, representing incoming traffic from secondary access roads (\emph{on-ramps}). Conversely, occupied bulk sites undergo detachment (desorption) at rate $\omega_D$, accounting for vehicles exiting toward secondary roads or parking facilities (\emph{off-ramps}).

Recent experiments have realized one-dimensional, single-file transport in autonomous active-matter platforms, including ring-shaped swarms of mobile robots \cite{Alonso2024robot}. Owing to hard-core exclusion and local, sensor-mediated speed control, these closed-loop systems exhibit self-organized stop-and-go waves and density-driven jamming transitions in the absence of particle exchange with the environment. By contrast, the TASEP-LK framework relaxes this strict conservation constraint, allowing us to analytically explore how open boundaries and bulk exchange reshape these non-equilibrium steady states.

\section{Second quantization formulation}
\label{sec:second}

The dynamics of continuous-time Markovian stochastic processes on discrete lattices are fundamentally governed by the chemical master equation,
\begin{equation}
    \frac{\partial}{\partial t}P(\eta,t) = \sum_{\eta'}\left\{T_{\eta'\rightarrow\eta}P(\eta',t) - T_{\eta\rightarrow\eta'}P(\eta,t)\right\},
\label{eq:master}
\end{equation}
which describes the temporal evolution of the probability distribution $P(\eta,t)$ over the configuration space $\eta(t) = \{n_1, n_2, \dots, n_k\}$ at time $t$. Here, $T_{\eta'\rightarrow\eta}$ denotes the transition rate from configuration $\eta'$ to $\eta$, and $n_j$ represents the particle occupation number at site $j$.

To convert this probabilistic description into an algebraic operator framework, we adopt Dirac’s bra-ket notation \cite{Dirac_1939} within the Fock space formalism \cite{Doi1, Doi2}. The state of the stochastic system at time $t$ is represented by a state vector $|\Psi(t)\rangle$, defined as a linear superposition over all accessible configurations $\ket{\eta} \equiv |n_1, n_2, \dots, n_k\rangle$:
\begin{equation}
    \ket{\Psi(t)}= \sum_{\eta} P(\eta,t) \ket{\eta}.
\label{eq:state_vector}
\end{equation}

We introduce local creation ($\hat{a}_j^\dagger$) and annihilation ($\hat{a}_j$) ladder operators acting on site $j$, which obey standard commutation relations:
\begin{equation}
    \left[\hat{a}_i, \hat{a}^\dagger_j\right] = \delta_{i,j},
    \quad \left[\hat{a}_i, \hat{a}_j\right] = \left[\hat{a}^\dagger_i, \hat{a}^\dagger_j\right] = 0,
\end{equation}
with the local vacuum state defined by $\hat{a}_j |0\rangle = 0$. A pure Fock state for a linear chain of $k$ sites is constructed from the vacuum state according to
\begin{equation}
    |\eta\rangle = |n_1, n_2, \dots, n_k\rangle = \prod_{j=1}^{k} \left(\hat{a}_j^\dagger\right)^{n_j}\ket{0}.
\label{eq:vecn}
\end{equation}
The operational actions of these second-quantized operators on an arbitrary occupation state follow
\begin{equation}
\begin{split}
    \hat{a}_j^\dagger |n_1, \dots, n_j, \dots\rangle &= |n_1, \dots, n_j+1, \dots\rangle, \\
    \hat{a}_j |n_1, \dots, n_j, \dots\rangle &= n_j |n_1, \dots, n_j-1, \dots\rangle.
\end{split}
\label{eq:opvec}
\end{equation}
To enforce hard-core exclusion, reflecting the physical constraint that each site can accommodate at most one particle ($n_j \in \{0, 1\}$), the local Hilbert space dimension at each site is restricted to two, imposing a nilpotency condition $(\hat{a}_j^\dagger)^2 = 0$. 

Within this algebraic framework, the master equation \eqref{eq:master} maps directly onto an imaginary-time Schrödinger-like equation ($\hbar=1$) \cite{Doi1, Doi2}:
\begin{equation}
    \frac{\partial}{\partial t} |\Psi(t)\rangle = -\hat{\mathcal{H}} |\Psi(t)\rangle,
\label{eq:sch}
\end{equation}
where $\hat{\mathcal{H}} \equiv \hat{\mathcal{H}}(\hat{a}_1^\dagger, \hat{a}_1, \dots, \hat{a}_k^\dagger, \hat{a}_k)$ is a non-Hermitian operator termed the quasi-Hamiltonian. The formal time evolution is governed by: 
\begin{equation}
    \left|\Psi(t)\right>=e^{-\hat{\mathcal{H}}t}\ket{\Psi(0)},
\label{eq:solsch}
\end{equation}
where $\left|\Psi(0)\right>$ specifies the initial configuration of the system. 

Following the algebraic formulation described in Ref.~\cite{Baez}, the quasi-Hamiltonian operator $\hat{\mathcal{H}}$ describing the combined TASEP-LK process is constructed as:
\begin{equation}
    \begin{split}
        \hat{\mathcal{H}} = & D\sum_{j=1}^{k-1}\left(\hat{a}_{j+1}^{\dagger}\hat{a}_j - \hat{a}^{\dagger}_j\hat{a}_j + \hat{a}_j^{\dagger}\hat{a}_{j}\hat{a}_{j+1}^{\dagger}\hat{a}_{j+1}\right) \\
        & + \omega_{A}\sum_{j=2}^{k-1}\left(1-\hat{a}_{j}\right)\hat{a}_{j}^{\dagger} + \alpha(1-\hat{a}_1)\hat{a}_1^{\dagger} \\
        & + \omega_{D}\sum_{j=2}^{k-1}\left(1-\hat{a}_j^{\dagger}\right)\hat{a}_j + \beta(1-\hat{a}_k^{\dagger})\hat{a}_k.
    \end{split}
    \label{eq:hamiltoniana}
\end{equation}

Notice that while the non-linear density products in $\hat{\mathcal{H}}$ correctly nullify the transition rates for blocked hops, it is the fundamental nilpotency algebra $(\hat{a}_j^\dagger)^2 = 0$ established earlier that strictly prevents the off-diagonal creation of multiply occupied states. Furthermore, in the context of second-quantized stochastic Liouvillians, Sato and Nishinari \cite{Sato2016relaxation} demonstrated that a non-unitary similarity transformation for Asymmetric Simple Exclusion Process (ASEP) with LK on periodic lattices can dynamically decouple fast bulk attachment-detachment modes from spatial hopping currents, yielding exact stationary states and system-size-independent relaxation gaps. In our open-boundary formulation, we solve for the steady state directly by constructing the zero-eigenvalue kernel of $\hat{\mathcal{H}}$.

The non-Hermitian quasi-Hamiltonian $\hat{\mathcal{H}}$ acts on a $2^k$-dimensional Fock space, isomorphic to a chain of $k$ spin-$1/2$ particles \cite{Heisenberg1928,Bethe1931,REIF, Pathria}. Its spectral decomposition relies on biorthonormal sets of right ($\ket{\psi_j}$) and left ($\bra{\phi_j}$) eigenvectors:
\begin{equation}
    \hat{\mathcal{H}} \ket{\psi_j} = \lambda_j \ket{\psi_j}, \quad \bra{\phi_j} \hat{\mathcal{H}} = \lambda_j \bra{\phi_j},
\label{eq:righteigen}
\end{equation}
normalized according to $\langle\phi_i|\psi_j\rangle = \delta_{ij}$. The formal time-evolution operator is thus expanded as
\begin{equation}
    e^{-\hat{\mathcal{H}}t} = \sum_j e^{-\lambda_j t} \ket{\psi_j}\bra{\phi_j}.
\end{equation}

\noindent In the field-theoretic Doi-Peliti formulation, physical expectation values at time $t$ are evaluated with respect to the left projection state $\bra{\psi_0} \equiv \langle 0| \exp\left(\sum_{j=1}^k \hat{a}_j\right)$, which satisfies $\bra{\psi_0} \hat{a}_j^\dagger = \bra{\psi_0}$. Conservation of total probability identically dictates that $\bra{\psi_0} \hat{\mathcal{H}} = 0$.

Evaluating the time derivative of the expectation value and utilizing this left-eigenstate probability conservation property, the temporal evolution of the mean occupancy $\langle n_j(t)\rangle$ is expressed via the commutator:
\begin{equation}
    \begin{split}
        \frac{d}{dt}\langle n_j \rangle &= -\langle \psi_0 | [\hat{n}_j, \hat{\mathcal{H}}] | \Psi(t) \rangle\\ &= -\langle \psi_0 | \hat{n}_j \hat{\mathcal{H}} | \Psi(t) \rangle.
    \end{split}
\end{equation}

\noindent The explicit evaluation of this commutator yields the exact discrete balance equations reflecting particle conservation. In the bulk ($j=2,\dots,k-1$), the rate of change of average density is given by
\begin{subequations}
\begin{equation}
\begin{split}
\frac{d\langle n_j\rangle}{dt}
&= \langle n_{j-1}(1-n_j)\rangle - \langle n_j(1-n_{j+1})\rangle \\
&\quad + \omega_A\bigl(1-\langle n_j\rangle\bigr) - \omega_D\langle n_j\rangle,
\end{split}
\end{equation}
while at the boundaries,
\begin{equation}
\begin{split}
\frac{d\langle n_1\rangle}{dt} &= \alpha\bigl(1-\langle n_1\rangle\bigr) - \langle n_1(1-n_2)\rangle,\\
\frac{d\langle n_k\rangle}{dt} &= \langle n_{k-1}(1-n_k)\rangle - \beta\langle n_k\rangle.
\end{split}
\label{eq:densidade}\end{equation}
\end{subequations}

\noindent Notice that these discrete balance equations couple the single-site expectation value $\langle n_j \rangle$ to the two-site joint expectation value $\langle n_j n_{j+1} \rangle$. These cross-terms encapsulate spatial correlations, which are rigorously quantified by the covariance of the occupation variables, $\mathrm{Cov}(n_i, n_j)$.

In the thermodynamic limit ($k \to \infty$, with continuous spatial coordinate $x = j/k \in [0,1]$ and lattice spacing $\epsilon = 1/k \to 0$), the standard MFA assumes spatial correlations are negligible ($\mathrm{Cov}(n_i, n_j) \approx 0$), yielding a continuous differential equation for the stationary density profile $\rho(x) = \langle n(x)\rangle$:
\begin{equation}
    \frac{\epsilon}{2}\,\partial_x^2 \rho + (2\rho - 1)\,\partial_x \rho + \Omega_A (1 - \rho) - \Omega_D \rho = 0,
\label{eq:parmeggiani2}
\end{equation}
where $\Omega_{A,D} = \omega_{A,D} k$ represent rescaled mesoscopic Langmuir rates. Subject to boundary conditions $\rho(0) = \alpha$ and $\rho(1) = 1-\beta$, Eq.~\eqref{eq:parmeggiani2} predicts sharp domain wall discontinuities. However, because MFA systematically discards these covariances, it fails to capture finite-size fluctuation smoothing and correlation-induced shifts, motivating an exact analytical approach.

In the asymptotic long-time limit ($t \to \infty$), the TASEP-LK dynamics project onto the unique zero eigenvalue $\lambda_1 = 0$ (with $\text{Re}(\lambda_j) > 0,\ \forall\ j > 1$), yielding
\begin{equation}
    \lim_{t\to\infty} e^{-\hat{\mathcal{H}}t} = |\Psi_{\text{s}}\rangle\langle\phi_1|,
\label{eq:steepest}
\end{equation}
where $|\Psi_{\text{s}}\rangle \equiv \ket{\psi_1}$ is the right eigenvector in the kernel of $\hat{\mathcal{H}}$, satisfying the stationary condition:
\begin{equation}
    \hat{\mathcal{H}} |\Psi_{\text{s}}\rangle = 0.
\label{eq:ss_condition}
\end{equation}
Since the left projection state $\bra{\psi_0}$ corresponds precisely to the left eigenvector $\bra{\phi_1}$ associated with $\lambda_1 = 0$, the exact expectation value of any observable $\hat{A}$ in the steady-state simplifies to:
\begin{equation}
    \langle A\rangle_{\text{s}} = \frac{\bra{\psi_0}\hat{A}\ket{\Psi_{\text{s}}}}{\bra{\psi_0}\Psi_{\text{s}}\rangle}.
\label{eq:valesperadoestacionario}
\end{equation}
Specifically, utilizing the projection identity, the local occupation operator $\hat{n}_j = \hat{a}_j^\dagger \hat{a}_j$ yields the exact stationary density profile:
\begin{equation}
    \langle n_{j}\rangle_{s} = \frac{\langle\psi_{0}|\hat{a}_{j}|\Psi_{s}\rangle}{\langle\psi_{0}|\Psi_{s}\rangle}.
\label{eq:density_exact}
\end{equation}

This analytical framework demonstrates that evaluating exact stationary observables entirely bypasses both explicit matrix diagonalization and time integration, reducing the problem to resolving the zero-eigenvalue kernel vector $|\Psi_{\text{s}}\rangle$ satisfying $\hat{\mathcal{H}} |\Psi_{\text{s}}\rangle = 0$ via the underlying Fock space algebra. Although the dimension of the configuration space grows exponentially as $2^k$, this exact operator methodology seamlessly handles mesoscopic chains up to $k \approx 16$ sites ($65,536$ basis states). Consequently, it provides a rigorous, non-perturbative benchmark for small-to-intermediate length scales, serving as a critical diagnostic tool to evaluate the breakdown of continuum mean-field approximations before finite-size artifacts overwhelm local density fluctuations.

%% file: results.tex
\section{Results}
\label{sec:results}

In this section, 
we present exact stationary-state analytical solutions for the stochastic process TASEP coupled with LK in a discrete linear chain of $k$ sites. To streamline the physical interpretation and eliminate parameter redundancies, we introduce two dimensionless control quantities: the boundary injection-to-extraction ratio, $r_b = \alpha/\beta$, and the bulk binding constant, $\kappa = \omega_A/\omega_D$, which dictates the relative balance between particle adsorption and desorption probabilities.

In accordance with representative benchmark parameters in the literature \cite{Derrida,parmegianni2003,parmegianni2004}, we fix $r_b = 1/3$ and $\kappa = 3$ throughout our analysis. To systematically explore the non-equilibrium phase space, we classify the stochastic transport dynamics into six distinct physical regimes based on the boundary injection strength $\alpha$ and the rescaled bulk desorption rate $\Omega_D = \omega_Dk$, as summarized in Table~\ref{tab:labels}. This analytical framework allows us to evaluate average density profiles $\langle n_j \rangle$, trace finite-size modifications, and quantify two-point spatial correlations without relying on MFA or suffering from stochastic sampling errors.

\begin{table}[htpb]\footnotesize
    \centering
    \caption{Nomenclature and parameter combinations for the six investigated physical transport regimes. All cases maintain a constant boundary ratio $r_b = \alpha/\beta = 1/3$ and bulk binding constant $\kappa = \omega_A/\omega_D = 3$.}
    \label{tab:labels}
    \begin{tabular}{c c c l}
    \hline\hline
    Label & $\alpha$ & $\Omega_D$ & Physical Transport Interpretation \\ \hline
    \textbf{LL} & 0.002 & 0.001 & Low injection, low bulk desorption \\ 
    \textbf{LC} & 0.002 & 0.100 & Low injection, competitive bulk kinetics \\    
    \textbf{ML} & 0.200 & 0.001 & Mid injection, low bulk desorption \\   
    \textbf{MC} & 0.200 & 0.100 & Mid injection, competitive bulk kinetics \\ 
    \textbf{MH} & 0.200 & 10.00 & Mid injection, high bulk desorption \\ 
    \textbf{HC} & 20.00 & 0.100 & High injection, competitive bulk kinetics \\ 
    \hline\hline
    \end{tabular}
\end{table}

\subsection{Stationary Density Profiles}

\begin{figure}[htpb]
    \centering
    \includegraphics[width=\columnwidth]{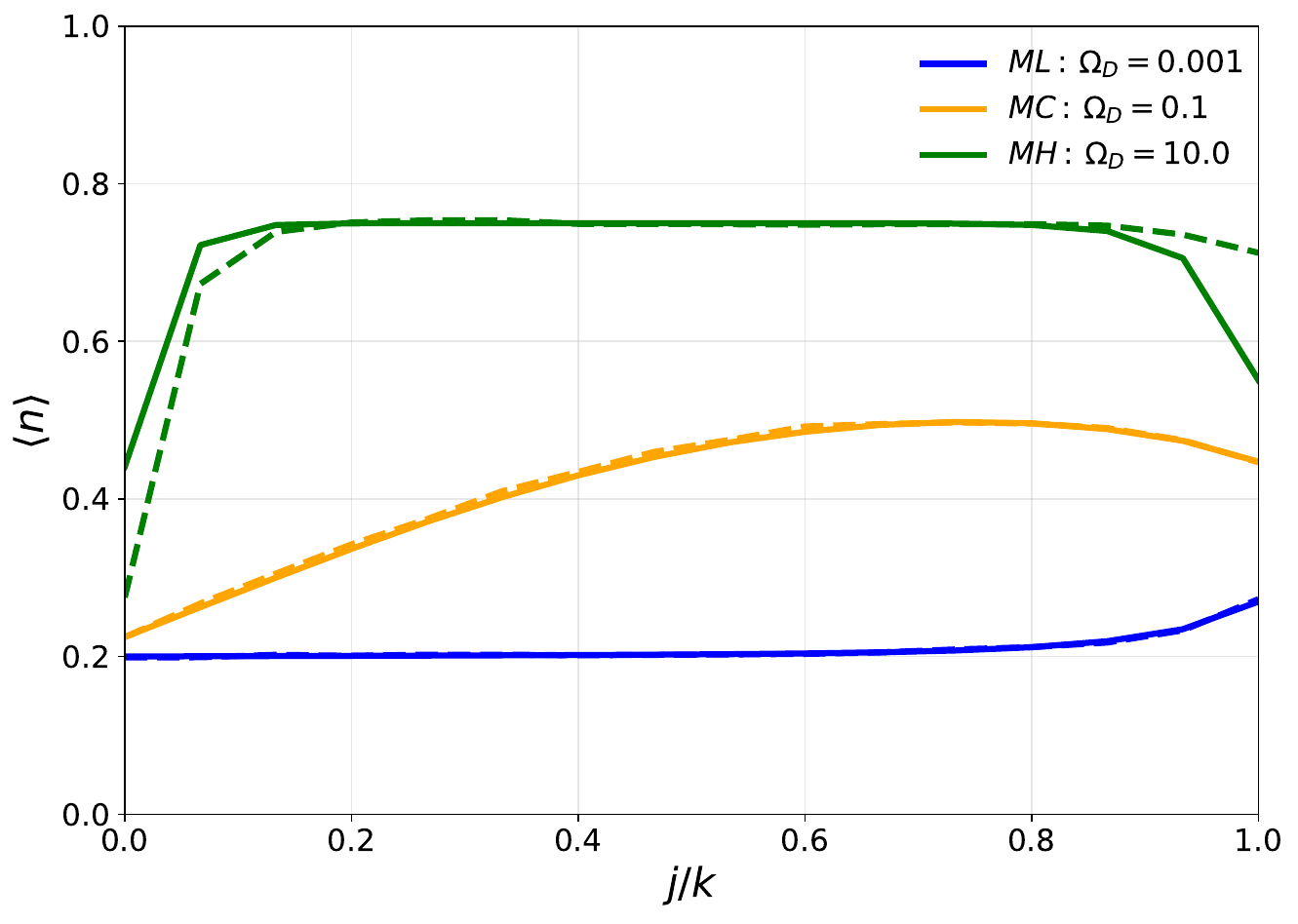}
    \caption{Benchmark comparison between exact calculations (solid lines) and stochastic Monte Carlo simulations (MCS, dashed lines) of the stationary density profile $\langle n_j\rangle$ as a function of the rescaled position $j/k$ on a lattice with $k=16$ sites, shown for low ($\Omega_D=0.001$, \textbf{ML}), competitive ($\Omega_D=0.1$, \textbf{MC}), and high ($\Omega_D=10$, \textbf{MH}) bulk exchange rates.}
    \label{fig:exact_vs_mcs-ML+MC+MH}
\end{figure}

We first analyze how varying the bulk desorption rate, $\Omega_D=\omega_Dk$, reshapes spatial density profiles under a fixed mid boundary injection, $\alpha = 0.2$ ($\beta = 0.6$). To validate our analytical framework and assess the numerical limits of stochastic sampling in mesoscopic chains ($k=16$), Fig.~\ref{fig:exact_vs_mcs-ML+MC+MH} provides a quantitative benchmark comparison between the proposed analytical formulation, and MCS implementations, represented by dotted curves, spanning regimes of low ($\Omega_D = 0.001$, ML), competitive ($\Omega_D = 0.1$, MC), and high ($\Omega_D = 10$, MH) bulk kinetics.

For high bulk exchange rates (\textbf{MH}, $\Omega_D = 10.0$), the microscopic rates per site in a mesoscopic chain of $k=16$ sites scale to $\omega_D = \Omega_D/k = 0.625$ and $\omega_A = \kappa\omega_D = 1.875$. Notice that the total local kinetic rate $(\omega_A + \omega_D = 2.50)$ exceeds the internal TASEP hopping rate ($D = 1$). Consequently, local adsorption-desorption dynamics occur on a faster timescale than longitudinal transport, forcing the local density across the bulk to lock rapidly onto the Langmuir equilibrium plateau, $\rho_{\text{eq}} = \kappa / (1 + \kappa) = 0.75$. Because boundary rates ($\alpha = 0.2, \beta = 0.6$) are lower than the internal adsorption rate ($\omega_A = 1.875$), boundary influences remain strictly confined to narrow boundary layers, manifesting as steep density drops to $\langle n_1 \rangle \approx 0.44$ at the entrance and $\langle n_k \rangle \approx 0.55$ at the exit. In traffic terms, the high density of secondary access roads (\emph{on-ramps}) saturates the highway, overriding boundary-driven flow. While MCS closely tracks the exact solution in the bulk, it exhibits slight deviations at boundary sites due to sampling stiffness under fast local exchange.

In the competitive regime (\textbf{MC}, $\Omega_D = 0.1$), microscopic bulk rates per site drop to $\omega_D = 0.00625$ and $\omega_A = 0.01875$, which are small compared to hopping ($D=1$). However, accumulated over $k=16$ sites, the net bulk exchange competes directly with boundary-driven transport, producing a smooth, broadly curved density profile. The density enters at $\langle n_1 \rangle \approx 0.225$, gradually rises to a peak of $\langle n_j \rangle \approx 0.50$ near $x \approx 0.75$ due to net bulk accretion, and drops slightly to $\langle n_k \rangle \approx 0.45$ near the exit. Because the chain length is mesoscopic, the integrated bulk kinetics are insufficient to fully enforce local equilibration, leaving the density profile well below the asymptotic Langmuir plateau ($\rho_{\text{eq}} = 0.75$).

Conversely, when bulk exchange is strongly suppressed (\textbf{ML}, $\Omega_D = 0.001 \implies \omega_D \approx 6.25 \times 10^{-5}$), unidirectional TASEP hopping dictates transport. The entry density strictly matches the boundary injection parameter, $\langle n_1 \rangle = \alpha = 0.20$. As vehicles progress along the main lane, weak bulk attachment induces a mild, monotonically increasing density trend across the lattice, reaching $\langle n_k \rangle \approx 0.27$ at the exit.

\begin{figure}[htpb]
    \centering
    \includegraphics[width=\columnwidth]{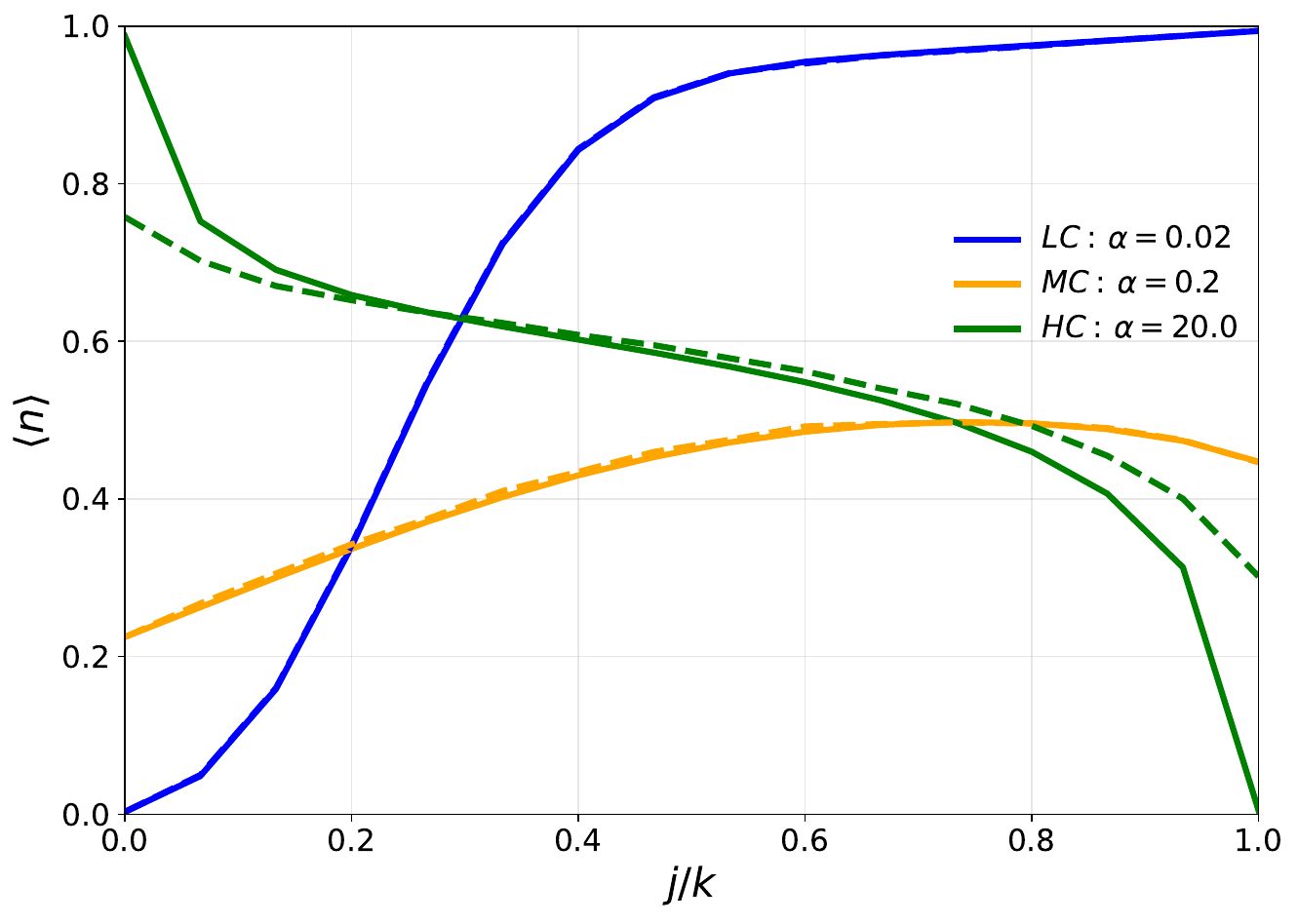}
    \caption{Exact stationary average density profiles $\langle n_j\rangle$ versus normalized lattice position $j/k$ for a $k=16$ site chain under competitive bulk kinetics ($\Omega_D = 0.1$). Curves illustrate the transition from low boundary drive and exit bottlenecking (\textbf{LC}, $\alpha = 0.002$), through intermediate drive (\textbf{MC}, $\alpha = 0.2$), to high boundary drive and rapid exit evacuation (\textbf{HC}, $\alpha = 20.0$). Solid lines denote exact Fock-space calculations, while dashed lines represent MCS estimates.}
    \label{fig:16s-exact_vs_mcs-LC+MC+HC}
\end{figure}

We next investigate the effect of boundary injection by varying $\alpha$ while keeping competitive bulk kinetics fixed at $\Omega_D = 0.1$. In analogy with the preceding analysis, Fig.~\ref{fig:16s-exact_vs_mcs-LC+MC+HC} compares the analytical predictions (solid curves) with MCS results (dashed curves) across low ($\alpha=0.002$, \textbf{LC}), mid ($\alpha = 0.2$, \textbf{MC}), and high ($\alpha=20$, \textbf{HC}) injection regimes. Although moderate rates (\textbf{LC} and \textbf{MC}) yield a highly accurate overlap between the curves obtained using the exact solution and the MCS, extreme driving ($\alpha=20, \beta=60$ in \textbf{HC}) exposes the relaxation limitations of MCS at the boundaries, whereas the exact operator solution rigorously resolves the steep boundary layer profiles.

In the \textbf{LC} regime ($\alpha = 0.002, \beta = 0.006$), boundary injection is negligible, resulting in a near-zero entry density ($\langle n_1 \rangle \approx 0$). As particles propagate downstream, continuous bulk adsorption (\emph{on-ramps}) progressively fills the lattice, causing a steep density growth toward $\rho_{\text{eq}} = 0.75$. Near the right boundary, the severely restricted extraction rate ($\beta = 0.006$) creates a major traffic bottleneck, leading to extreme particle accumulation and forcing complete boundary saturation, $\langle n_k \rangle \approx 1.0$.

Conversely, the \textbf{HC} regime ($\alpha = 20.0, \beta = 60.0$) demonstrates an inverted physical symmetry relative to \textbf{LC}. Massive boundary injection saturates the entrance ($\langle n_1 \rangle \approx 1.0$). Downstream, bulk desorption (\emph{off-ramps}) acts as a relief mechanism, continuously draining particles and pulling the density profile down toward $\rho_{\text{eq}} = 0.75$. At the right boundary, the extremely high extraction rate ($\beta = 60.0$) acts as an unobstructed exit, completely evacuating the final site ($\langle n_k \rangle \approx 0.0$). Finally, the behavior of the \textbf{MC} regime remains consistent with that reported in Fig.~\ref{fig:exact_vs_mcs-ML+MC+MH}.

\subsection{Thermodynamic Limit and Domain Wall Formation}
\label{subsec:dw_formation}

The macroscopic stationary behavior of the hybrid TASEP-LK process in the continuum thermodynamic limit ($k \to \infty$) is fundamentally characterized by the emergence of localized spatial shock waves, or Domain Walls (DWs) \cite{parmegianni2003,parmegianni2004}. While continuum theories predict sharp mathematical discontinuities in the density profile, physical transport in finite mesoscopic systems exhibits smooth, continuous spatial gradients.

\begin{figure}[htpb]
    \centering
    \includegraphics[width=\columnwidth]{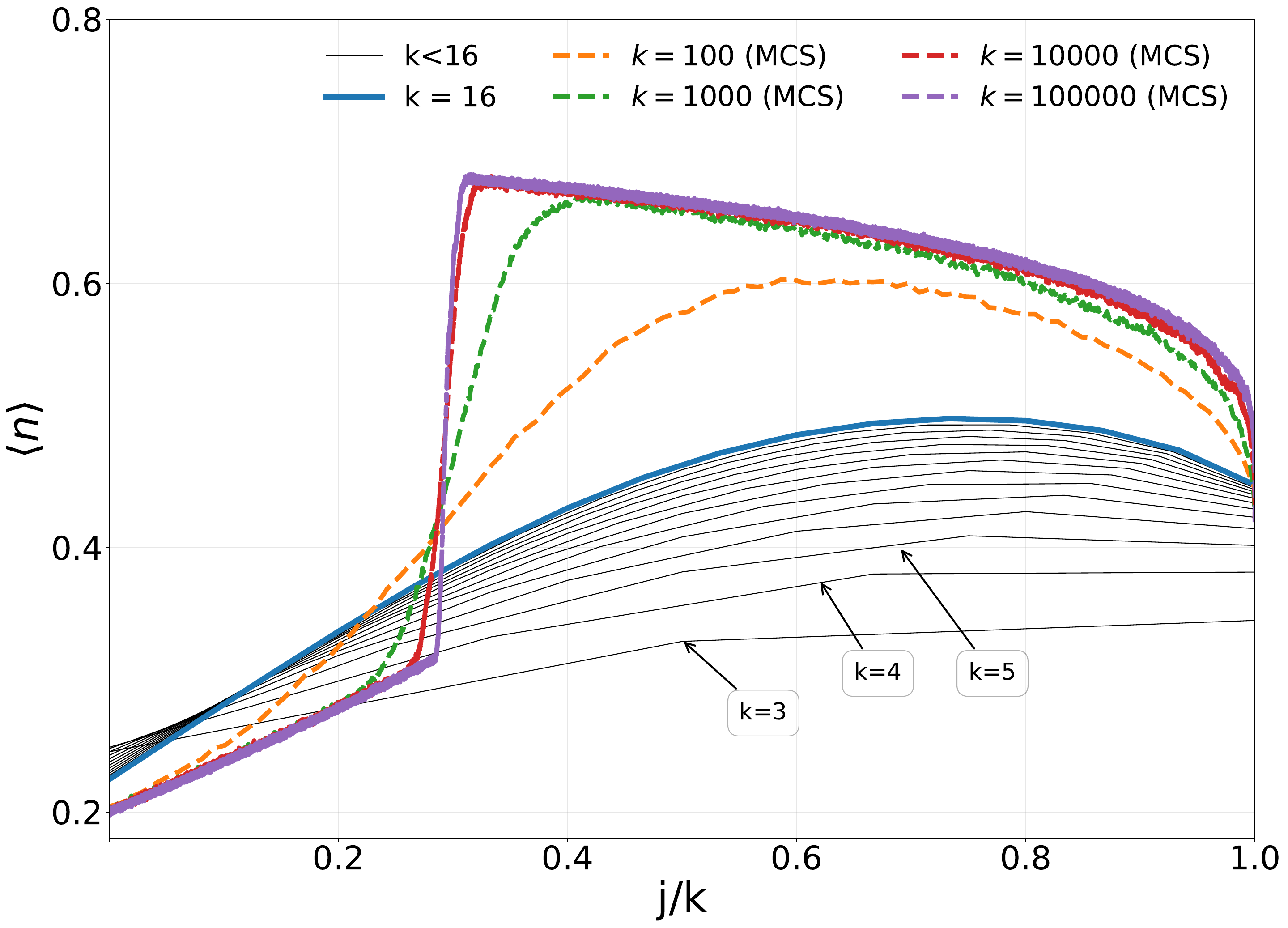}
    \caption{Transition from mesoscopic fluctuation smoothing to macroscopic domain wall formation in the \textbf{MC} regime. Exact Fock-space solutions for $k < 16$ (black solid curves) and $k = 16$ (solid blue curve) are benchmarked against MCS results (dashed lines) for macroscopic chains ($k = 100$ to $k = 100,000$). As $k$ increases, the smooth mesoscopic gradient continuously sharpens into a localized DW centered at $x_{\text{DW}} \approx 0.28$.}
    \label{fig:exact_vs_mcs-variable}
\end{figure}

This discrepancy stems from finite-size fluctuation smoothing: in short lattices, statistical fluctuations in particle hopping and bulk exchange smear out the spatial position of the shock front across the chain length. To map the transition from mesoscopic smoothing to continuum shock formation, we compute exact stationary density profiles using our Fock space operator framework for chain sizes ranging from $k=3$ to $k=16$ under the competitive kinetics regime (\textbf{MC}, $\Omega_D = 0.1$, $\alpha = 0.2$). To bridge these exact mesoscopic results with macroscopic behavior, we extend the system size up to $k = 100,000$ using stochastic Monte Carlo simulations (MCS), as illustrated in Fig.~\ref{fig:exact_vs_mcs-variable}.

For very short chains ($k \le 5$), boundary injection and extraction constraints dominate the dynamics, limiting the maximum bulk density below $0.45$. As the chain length increases to $k=16$, continuous particle accretion along bulk \emph{on-ramps} elevates the central density toward the Langmuir equilibrium level, producing a broad, smooth profile. When the lattice size is scaled further ($k = 100 \to 100,000$), the integrated bulk exchange progressively dominates over longitudinal boundary diffusion. Consequently, the characteristic shock width narrows relative to the total lattice length, sharpening the smooth mesoscopic gradient into a steep domain wall. For $k \ge 10,000$, the discrete profile converges onto a sharp discontinuity at $x_{\text{DW}} \approx 0.28$, separating a low-density region ($\rho < 0.35$) from a high-density downstream plateau ($\rho \approx 0.68$). Crucially, this finite-size scaling demonstrates that domain-wall structures cannot develop in small mesoscopic systems ($k \le 16$), emerging distinctly only when $k \ge 100$.

Through the lens of traffic flow engineering, this finite-size transition carries a clear physical interpretation: short highway segments ($k \le 16$) efficiently buffer incoming traffic from secondary access roads (\emph{on-ramps}) due to boundary-driven fluctuation dissipation, maintaining smooth vehicular flow without localized bottlenecks. However, along extended highways ($k \ge 1,000$), cumulative vehicle entries from bulk \emph{on-ramps} eventually saturate downstream carrying capacity. This localized capacity overflow triggers a sharp traffic shock wave—the Domain Wall—marking an abrupt transition from free-flowing vehicular movement to dense, congested conditions.

\subsection{Global Phase Diagram and Parametric Mapping}
\label{subsec:parametric_scan}

\begin{figure}[htpb]
    \centering
    \includegraphics[width=\columnwidth]{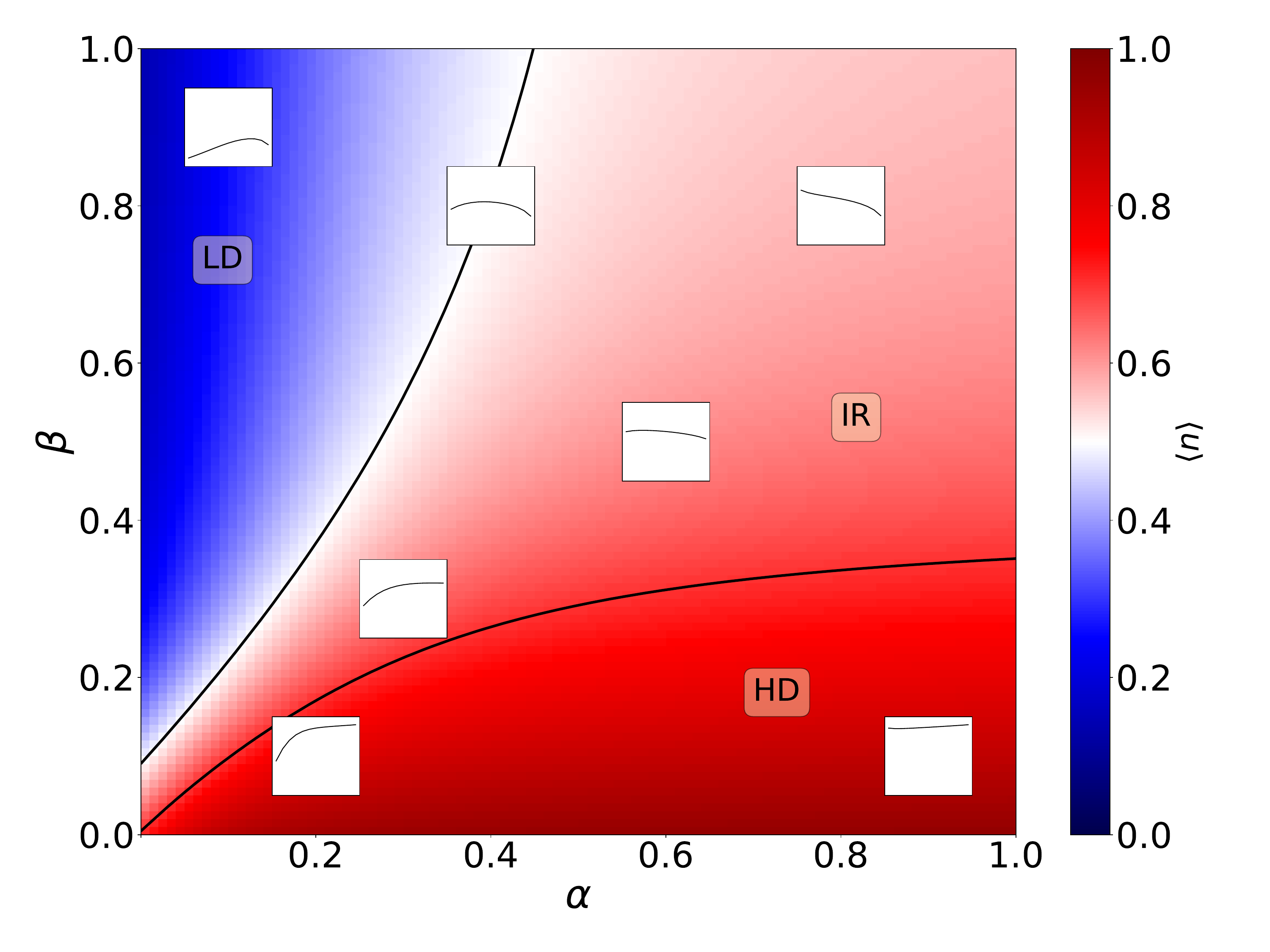}
    \caption{Exact parametric phase diagram mapping the average stationary density $\bar{\rho}$ over the boundary rate space $(\alpha, \beta) \in [0,1] \times [0,1]$ for a $k=13$ site chain with competitive bulk exchange ($\Omega_D = 0.1$, $\kappa = 3$). The color map encodes $\bar{\rho}$, with solid black contour lines marking the boundaries $\bar{\rho} = 0.5$ and $\bar{\rho} = 0.7$. Insets display representative local density profiles $\langle n_j \rangle$ vs $j/k$ across the low-density (\textbf{LD}), high-density (\textbf{HD}), and intermediate region (\textbf{IR}) regimes.}
    \label{fig:13s-parametric_scan_with_inset}
\end{figure}

To synthesize the global stationary behavior of the TASEP-LK system, we construct a continuous two-dimensional phase diagram mapping the spatial average lattice density $\bar{\rho} = \frac{1}{k} \sum_{j=1}^k \langle n_j \rangle$ across the boundary control parameter space $(\alpha, \beta) \in [0, 1] \times [0, 1]$. Figure~\ref{fig:13s-parametric_scan_with_inset} depicts the parametric scan evaluated under fixed competitive bulk exchange conditions ($\Omega_D = 0.1 \therefore \omega_D \approx 0.008$, $\kappa = 3$), with insets illustrating representative local density profiles $\langle n_j \rangle$ across distinct topological regions.

To achieve a high-resolution continuous rasterization of the phase space, we performed a large number of exact calculations over a dense grid ($100\times100$) of boundary rate pairs for a mesoscopic chain of $k=13$ lattice sites. Provides an exact, fluctuation-rigorous map of the stationary phase landscape without stochastic sampling noise or mean-field decoupling approximations.

In mesoscopic chains ($k=13$), finite-size fluctuations smooth out sharp phase boundaries, preventing the strict thermodynamic phase coexistence characterized by singular domain walls. Consequently, we partition the parametric space into three operational transport regimes demarcated by explicit average density thresholds:
\begin{equation*}\footnotesize
    \text{Transport Regime} = 
    \begin{cases}
    \text{Low-Density (\textbf{LD})}, & \bar{\rho} < 0.5, \\
    \text{Intermediate Region (\textbf{IR})}, & 0.5 < \bar{\rho} \le 0.7, \\
    \text{High-Density (\textbf{HD})}, & \bar{\rho} \ge 0.7.
    \end{cases}
\end{equation*}
In Fig.~\ref{fig:13s-parametric_scan_with_inset}, solid black contour lines represent the explicit isolines $\bar{\rho} = 0.5$ and $\bar{\rho} = 0.7$ bounding these regimes.

\begin{itemize}
    \item \textbf{Low-Density (\textbf{LD}) Phase ($\bar{\rho} < 0.5$):} Dominated by low boundary injection rates ($\alpha < \beta$), this regime maintains a sparse lattice occupancy. As shown in the top-left inset, the density profile remains low throughout the chain ($\langle n_j \rangle < 0.5$). The average occupancy can be finely tuned by varying $\alpha$ within a restricted window ($0 \le \alpha \le 0.45$ for $\beta \ge 0.09$). In traffic engineering terms, \textbf{LD} corresponds to a light, free-flowing traffic corridor where low entry demand prevents congestion bottlenecks, even under modest exit rates.

    \item \textbf{High-Density (\textbf{HD}) Phase ($\bar{\rho} \ge 0.7$):} Governed by low extraction rates ($\beta < \alpha$), this regime forces severe particle accumulation. The boundary extraction parameter $\beta$ operates within a narrow window ($\beta \le 0.35$); exceeding this threshold transitions the system into lower density regimes. In this state, downstream exit bottlenecks cause extensive congestion, driving local site occupancies near total saturation ($\langle n_j \rangle \to 1.0$), as depicted in the lower insets.

    \item \textbf{Intermediate Region (\textbf{IR}) ($0.5 < \bar{\rho} \le 0.7$):} Bounded between the $\bar{\rho} = 0.5$ and $\bar{\rho} = 0.7$ isolines, the \textbf{IR} represents a flexible transport domain characterized by nearly uniform, moderate local occupation profiles. Here, wide variations in both $\alpha$ and $\beta$ permit continuous tuning of $\bar{\rho}$ without triggering localized capacity collapse or complete road starvation. From a traffic management perspective, the \textbf{IR} models a high-throughput corridor operating at optimal capacity—carrying substantial vehicular flow without suffering from sudden shock-wave jams.
\end{itemize}

\begin{figure}[htpb]
    \centering
    \includegraphics[width=\columnwidth]{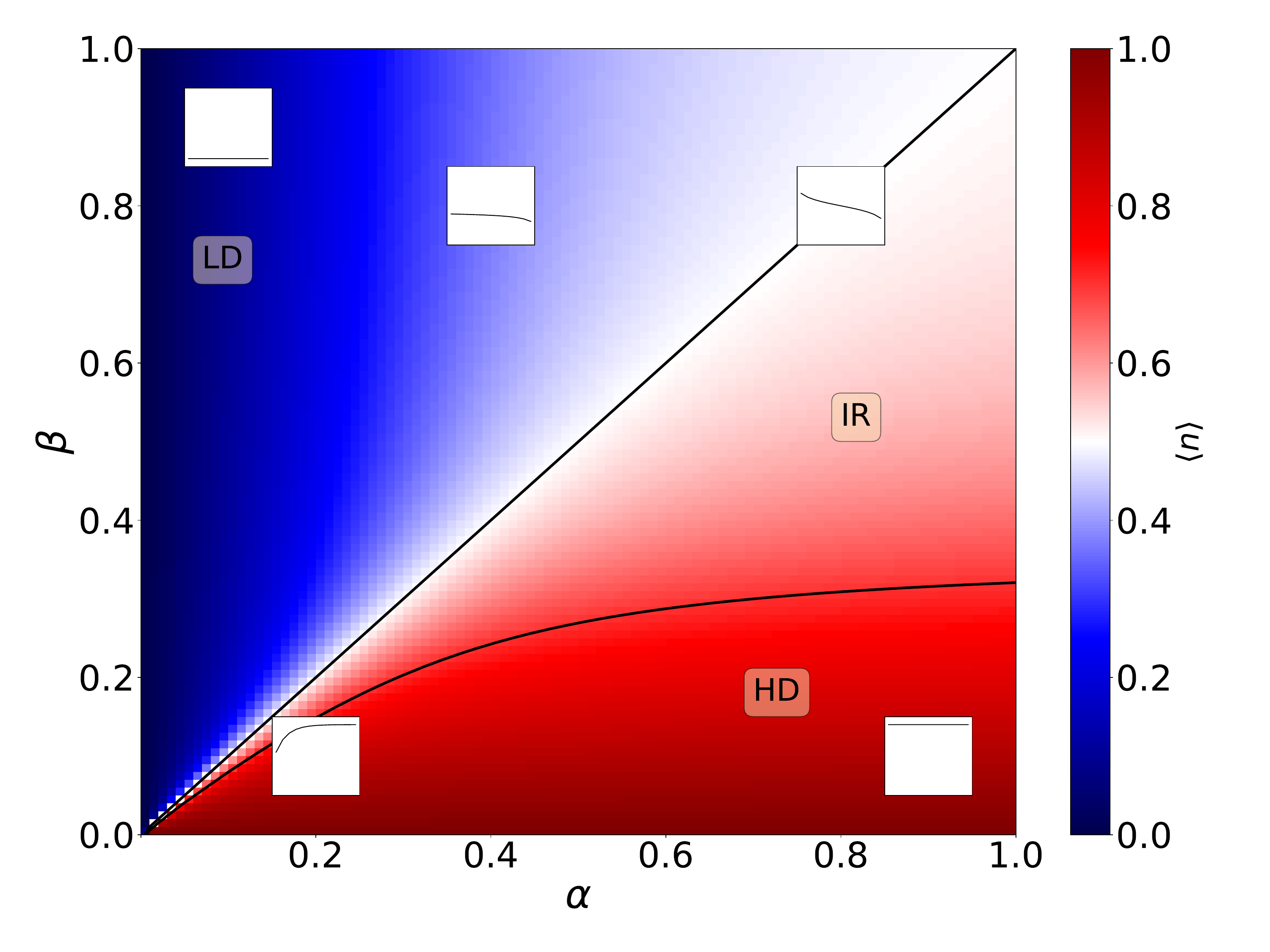}
    \caption{Exact parametric phase diagram of the stationary mean density $\bar{\rho}$ in the boundary plane $(\alpha,\beta) \in [0, 1] \times [0, 1]$ for a pure TASEP chain of length $k=13$ ($\Omega_D = \Omega_A = 0$). Solid contour lines represent the thresholds $\bar{\rho} = 0.5$ and $\bar{\rho} = 0.7$, delimiting the \textbf{LD}, \textbf{IR}, and \textbf{HD} regimes. Insets display representative local density profiles $\langle n_j \rangle$ as a function of $ j/k$.}
    \label{fig:13s-parametric_scan_tasep}
\end{figure}

To isolate the precise influence of bulk adsorption-desorption kinetics on mesoscopic phase topology, we computed a benchmark parametric scan for a pure TASEP chain without Langmuir kinetics ($\Omega_D = \Omega_A = 0$), as shown in Fig.~\ref{fig:13s-parametric_scan_tasep}.

\noindent Comparing Fig.~\ref{fig:13s-parametric_scan_with_inset} (TASEP-LK) and Fig.~\ref{fig:13s-parametric_scan_tasep} (pure TASEP) highlights several fundamental physical transformations induced by bulk kinetics in mesoscopic systems:
\begin{enumerate}
    \item \textbf{Expansion of the Moderate-Density Domain:} Coupling Langmuir kinetics substantially expands the parameter area occupied by the Intermediate Region (\textbf{IR}). In pure TASEP, intermediate occupancies are constrained to a narrow diagonal band near $\alpha \approx \beta$. In TASEP-LK, net bulk adsorption ($\kappa = 3$) inflates the \textbf{IR} region across a much wider range of boundary rates, offering broader operational flexibility to maintain moderate traffic loads.
    
    \item \textbf{Parametric Asymmetry and Exit Decoupling:} In pure TASEP, the phase topology exhibits symmetric behavior along the main diagonal $\alpha = \beta$. In contrast, bulk Langmuir kinetics introduces a strong topological asymmetry: the upper contour line ($\bar{\rho} = 0.7$) shifts downward toward lower $\beta$ values ($\beta \approx 0.35$). Consequently, bulk adsorption acts as a secondary particle source that elevates internal density, allowing the system to maintain moderate bulk throughput over a wider spectrum of entrance rates $\alpha$, largely independent of exit bottlenecks.
\end{enumerate}

\subsection{Spatial Correlations and Breakdown of Mean-Field Theory}
\label{subsec:spatial_correlations}

Beyond single-site stationary density profiles, the exact operator approach provides direct analytical access to higher-order joint spatial correlation functions without resorting to decoupling approximations. To quantify two-point spatial correlations and evaluate the breakdown of the factorized mean-field assumption ($\langle n_i n_j \rangle \approx \langle n_i \rangle \langle n_j \rangle$), we compute the spatial covariance matrix:
\begin{equation}
    \mathrm{Cov}(n_i,n_j) = \langle n_i n_j \rangle - \langle n_i \rangle \langle n_j \rangle.
    \label{eq:covariance_def}
\end{equation}

A comprehensive parametric investigation (not shown here) across all kinetic regimes reveals that inter-site spatial correlations are most strongly amplified in the low-rate \textbf{LL} regime ($\alpha = 0.002, \Omega_D = 0.001 \therefore \omega_D \approx 6.25 \times 10^{-5}$). In this slow-kinetics domain, the low frequency of boundary injection and bulk exchange events allows local density fluctuations to persist over extended relaxation timescales. Consequently, particle transport becomes highly collective, generating strong spatial memory that fundamentally invalidates mean-field decoupling. In contrast, regimes with faster exchange kinetics (such as \textbf{MC}) exhibit rapid dynamic mixing, which suppresses off-diagonal spatial covariances to negligible values ($\mathrm{Cov}(n_i,n_j) \approx 0.01$).

\begin{figure}[htpb]
    \centering
    \includegraphics[width=\columnwidth]{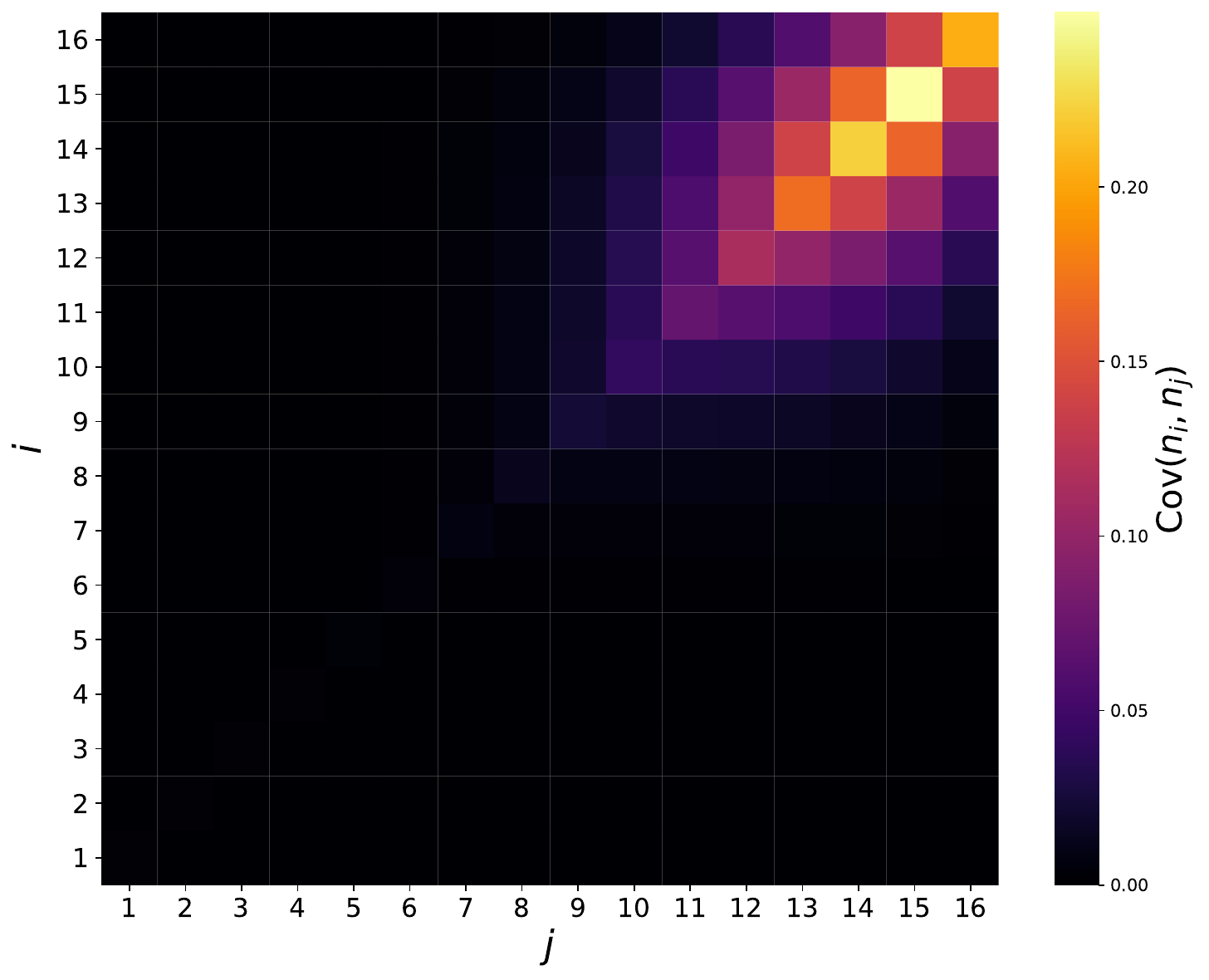}
    \caption{Exact stationary covariance matrix for a $k=16$ site chain in the low-rate \textbf{LL} regime ($\alpha = 0.002, \Omega_D = 0.001$). Diagonal entries represent local site variances $\text{Var}(n_i)$, while non-zero off-diagonal elements reveal strong spatial correlations concentrated exclusively in the downstream lattice region ($i, j \ge 9$).}
    \label{fig:16s-LL-joint_cumulants_heatmap}
\end{figure}

Fig. \ref{fig:16s-LL-joint_cumulants_heatmap} displays the full $16 \times 16$ covariance matrix for a $k=16$ chain in the \textbf{LL} regime. The diagonal elements correspond to local occupation variances, $\text{Var}(n_i)=\langle n_i \rangle (1-\langle n_i \rangle)$, which reach a maximum at the penultimate site ($\text{Var}(n_{15}) \approx 0.25$) rather than the final boundary. Because the extraction rate is extremely small ($\beta=0.006$), the right boundary effectively behaves as a nearly closed gate; the terminal site remains almost always occupied, strongly suppressing its local variance. Consequently, significant non-zero off-diagonal correlations develop exclusively in the adjacent downstream segment ($i, j \ge 9$), with the strongest nearest-neighbor spatial correlation occurring between sites 14 and 15 ($\mathrm{Cov}(n_{14}, n_{15}) \approx 0.16$). This feature persists with increasing chain length, consistent with a DW pinned at the boundary whose fluctuations propagate upstream. In contrast, mean-field theory cannot capture this spatially displaced fluctuation structure.

\begin{figure}[htpb]
    \centering
    \includegraphics[width=\columnwidth]{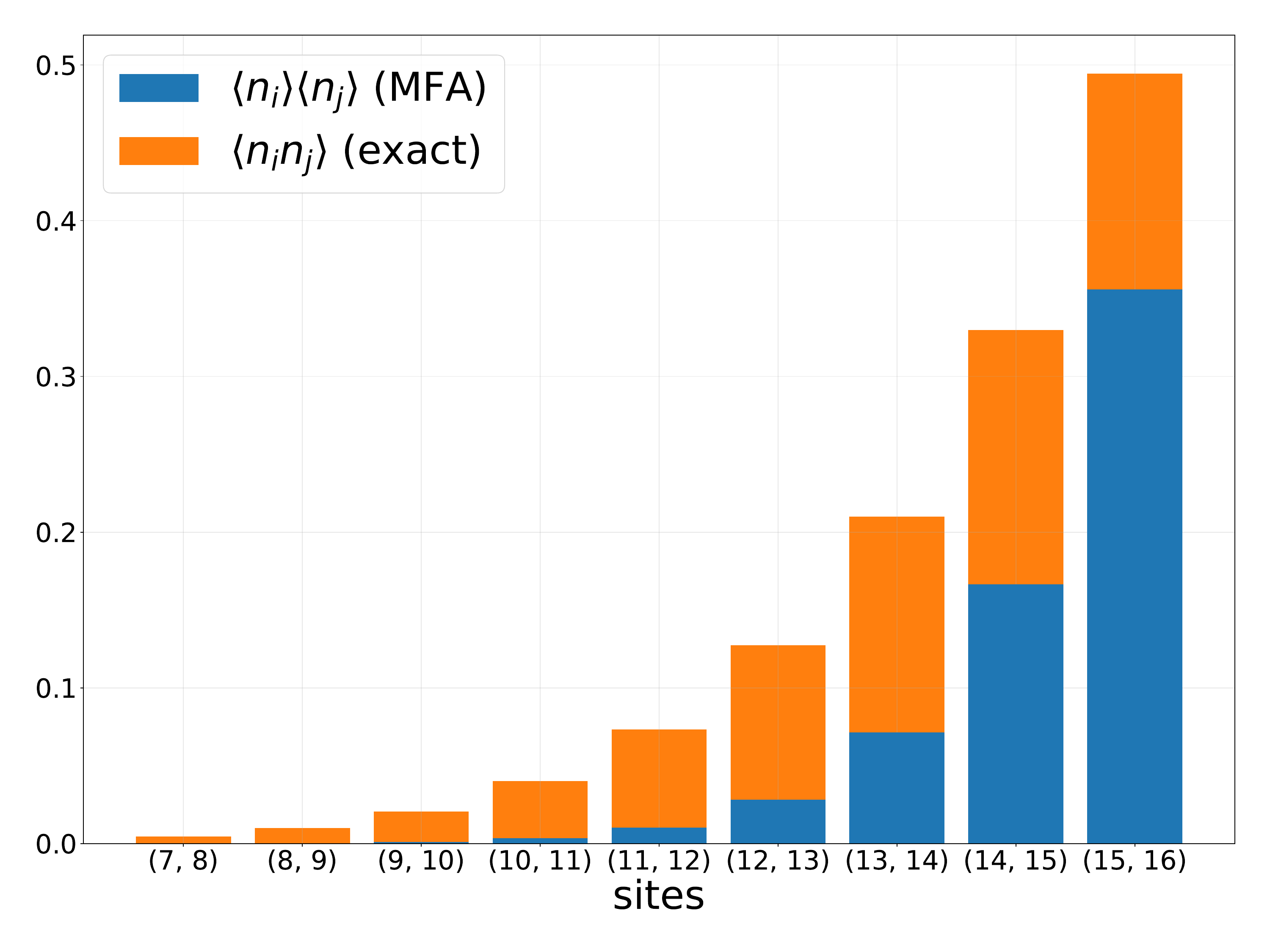}
    \caption{Direct comparison between exact stationary joint occupation moments $\langle n_i n_j \rangle$ (orange bars) and factorized mean-field predictions $\langle n_i \rangle \langle n_j \rangle$ (blue bars) for adjacent downstream site pairs in a $k=16$ chain under the \textbf{LL} regime. Mean-field theory severely underestimates joint occupation probabilities as spatial correlations intensify downstream.}
    \label{fig:16s-LL-exact_vs_meanfield}
\end{figure}

To explicitly demonstrate the failure of mean-field theory in this mesoscopic regime, Fig.~\ref{fig:16s-LL-exact_vs_meanfield} contrasts the exact joint occupation moments $\langle n_i n_j \rangle_{\text{exact}}$ against their factorized mean-field counterparts $\langle n_i \rangle \langle n_j \rangle_{\text{MFA}}$ for adjacent downstream site pairs $(j, j+1)$. We observe that the mean-field approximation (MFA) systematically and severely underestimates the true joint occupancy probabilities across the entire downstream region. For instance, for the adjacent pair $(14, 15)$, the exact joint moment reaches $\langle n_{14} n_{15} \rangle \approx 0.33$, whereas the factorized mean-field product predicts only $\langle n_{14} \rangle \langle n_{15} \rangle \approx 0.17$—a striking twofold underestimation.

Because the spatial covariances are strictly positive ($\mathrm{Cov}(n_i, n_j) > 0$), particles in the slow-kinetics regime exhibit effective spatial clustering: the presence of a particle at site $i$ substantially increases the conditional probability that adjacent downstream sites are also occupied. By discarding these many-body spatial correlations, mean-field theory fails to capture the cooperative nature of non-equilibrium transport, underscoring the necessity of exact methods for low-dimensional stochastic systems.

%% file: conclusion.tex
\section{Conclusion}
\label{sec:conclusion}

In this work, we employ a field-theoretic operator formulation in Fock space to evaluate steady-state transport properties, global phase diagrams, and spatial correlation functions in finite driven-diffusive systems, focusing on the totally asymmetric simple exclusion process coupled with Langmuir kinetics (TASEP-LK). By constructing biorthonormal state eigenvectors within the Doi-Peliti framework, we established a representation that reduces the evaluation of stationary observables to solving the sparse matrix kernel equation. This operator approach completely circumvents the severe $\mathcal{O}(2^{3k})$ computational complexity associated with full matrix diagonalization and time-dependent matrix exponentiation.

Applying this exact methodology to mesoscopic chains of up to $k = 16$ lattice sites, we systematically investigated six fundamental transport regimes spanning low, competitive, and high rates of boundary injection ($\alpha$) and bulk exchange ($\Omega_D$). Our exact calculations demonstrate that finite-size stochastic fluctuations naturally smooth out continuous field-theoretic singularities: no sharp domain walls emerge in short-to-mesoscopic chains, yielding continuous, broadly curved density profiles across all boundary and bulk kinetic conditions.

Through finite-size scaling ($k = 3$ to $100,000$), we tracked the crossover toward the continuous thermodynamic limit ($k \to \infty$). As system size increases, ongoing bulk adsorption along \emph{on-ramps} progressively elevates the central density toward the Langmuir equilibrium plateau ($\rho_{\text{eq}} = 0.75$), while the spatial density gradient steepens rapidly near $x \approx 0.28$. This confirms that singular Domain Walls represent mathematical idealizations valid strictly in the thermodynamic continuum limit, whereas mesoscopic transport segments handle spatial fluctuations smoothly.

Exact phase-diagram calculations for chains with $k=13$ sites resolve three distinct stationary density regimes. Incorporating Langmuir kinetics significantly enlarges the intermediate-density region in parameter space, enabling its realization over a much broader range of boundary injection ($\alpha$) and extraction ($\beta$) rates than is possible in the pure TASEP limit.

Crucially, our evaluation of spatial covariances ($\mathrm{Cov}(n_i, n_j)$) exposes the fundamental breakdown of standard factorized Mean-Field Approximations (MFA). In the slow-kinetics regime (\textbf{LL}, $\alpha = 0.002, \Omega_D = 0.001$), persistent density fluctuations amplify spatial correlations, peaking at $\mathrm{Cov}(n_{14}, n_{15}) \approx 0.16$. Consequently, MFA severely underestimates joint occupancy probabilities, predicting a factorized product of $\langle n_{14} \rangle \langle n_{15} \rangle \approx 0.17$, whereas the exact joint moment reaches $\langle n_{14} n_{15} \rangle \approx 0.33$, a two-fold error.

These results demonstrate that while continuum field theories provide valuable global phase boundaries, exact algebraic methods are indispensable for modeling finite non-equilibrium transport systems. Prime examples include vehicular traffic flow along short highway segments and urban merge zones \cite{CHOWDHURY2000199, Schadschneider_1997, Wassmer2026}, as well as the intracellular transport of biological molecular motors along microtubule networks \cite{parmegianni2003, parmegianni2004, Bressloff}. Furthermore, while standard TASEP exhibits a strictly monotonic decrease of particle velocity with density, driven exclusion processes coupled to auxiliary environmental fields, such as cellular-automaton models for ant trails governed by dynamic pheromone deposition \cite{Debashish_Chowdhury_2002,Nishinari_2003, Ngo2024ant}, can display non-monotonic flow-density relations. In traffic engineering terms, continuous mean-field models artificially overpredict localized traffic breakdowns on short road segments; in reality, mesoscopic stochastic fluctuations smooth out potential bottlenecks created by secondary access ramps.

Future extensions of our analytical methodology could incorporate spatial heterogeneity (e.g., localized bottlenecks or variable speed limits), multi-lane coupling dynamics, or non-Markovian headway memory effects in vehicular queues. Furthermore, extending the operator formalism to evaluate first-passage and exit time distributions \cite{redner, gerson,Szabo1980Firstpassage, arthur_artigo,Nicolau2021MFP} will be a critical next step in characterizing the transport efficiency, dynamical persistence, and kinetic relaxation of these mesoscopic systems.